\documentclass[final,english]{bullsrsl}

\usepackage[latin1]{inputenc}
\usepackage[T1]{fontenc}
\usepackage{amssymb}
\usepackage{amsmath}
\def\medd{\dot{M}_{\rm Edd}}

\usepackage{natbib} 

\usepackage{graphicx}
\usepackage{hyperref}
\hypersetup{colorlinks=true, linkcolor=blue, filecolor=blue, urlcolor=blue,citecolor=blue}

\def\rg{r_{\rm g}}

\begin{document}
\title{Outflow Behavior from the Transonic Advective Disks: A Hydrodynamical Simulation Study}



  \author[affil={1,2}, corresponding]{Sanjit}{Debnath}
  \author[affil={1}]{Indranil}{Chattopadhyay}
 \author[affil={3}]{Raj Kishor} {Joshi}
 \author[affil={4}]{Philippe} {Laurent}
 \author[affil={1,2}]{Priyesh Kumar} {Tripathi}
 \author[affil={2}]{M. Saleem}{Khan}

\affiliation[1]{Aryabhatta Research Institute of Observational Sciences, Manora Peak, Nainital, 263001, India}
\affiliation[2]{Department of Applied Physics, Mahatma Jyotiba Phule Rohilkhand University, Bareilly, Uttar Pradesh, 243006, India}
\affiliation[3]{Nicolaus Copernicus Astronomical Center, Polish Academy of Sciences, Bartycka 18, PL-00-716, Warsaw, Poland}
\affiliation[4]{IRFU / Service d'Astrophysique, Bat. 709 Orme des Merisiers, CEA Saclay, 91191 Gif-sur-Yvette, Cedex France}

\correspondance{sanjitphysics1998@gmail.com, sdebnath@aries.res.in}


\maketitle

\begin{abstract}
We investigate the properties of outflows from the transonic advective accretion disk using hydrodynamical numerical simulations. We consider two different disk temperatures with an order-of-magnitude difference. For the hotter disk, we adopt initial conditions for velocity, specific angular momentum, and temperature from analytical solutions. In the colder disk case, the velocity and angular momentum profiles are kept identical, while an order of magnitude reduction in the temperature. The simulations are performed in the presence of viscosity and radiative cooling, considering bremsstrahlung and synchrotron processes. In both disk models, the outflow rate increases with viscosity. We also examine the poloidal velocity structures for both cases. We analyze the influence of viscosity on the mass flux-weighted energy and momentum fluxes of the outflows. Our results show that both energy and momentum fluxes increase with higher viscosity and may play a significant role in accretion feedback mechanisms.
\end{abstract}





\section{Introduction}

Accretion flows in which advective transport plays a significant role have emerged as a robust framework for interpreting the high-energy emission observed from black hole systems. In particular, the advection-dominated accretion flow (ADAF) paradigm \citep{1977ApJ...214..840I,1995ApJ...452..710N} and the global transonic advective solutions \citep{1987PASJ...39..309F,1989ApJ...347..365C} successfully describe hot, optically thin flows capable of producing X-ray radiation. Unlike the spherically symmetric Bondi accretion solution \citep{1952MNRAS.112..195B}, rotating advective flows are characterized by the possibility of multiple sonic points. Under appropriate conditions of angular momentum and energy, such multi-transonic solutions can host stationary shock transitions \citep{1987PASJ...39..309F,1989ApJ...347..365C,2009ApJ...702..649D,2011IJMPD..20.1597C,2013MNRAS.430..386K,2014MNRAS.443.3444K}. 
The post-shock region of a sub-Keplerian accretion flow was proposed to correspond to the hot coronal component responsible for inverse Comptonization \citep{1995ApJ...455..623C}. In this framework, the compressed and heated post-shock plasma acts as an efficient source of hard X-ray photons. 
Using time-dependent hydrodynamic simulations, several studies demonstrated both the stability and oscillatory behavior of shock fronts under different physical conditions \citep{1993ApJ...417..671C,1994ApJ...425..161M,1996ApJ...457..805M,1996ApJ...470..460M,2004A&A...421....1C}. More recently, numerous computational works have explored inviscid advective transonic flows to interpret a variety of accretion-driven phenomena around black hole candidates \citep{2010MNRAS.403..516G,2012MNRAS.425.2413O,2012ApJ...758..114G,2014MNRAS.437.1329G,2017MNRAS.472.4327S,2017MNRAS.472..542K,2019MNRAS.482.3636K,2020ApJ...904...21P,2022MNRAS.514.5074O,2023MNRAS.519.4550G,2025ApJ...990...35D,2026Univ...12...77J}.

Previous numerical simulations \citep{1998MNRAS.299..799L,2012MNRAS.421..666G,2013MNRAS.430.2836G,2014MNRAS.442..251D,2015MNRAS.453..147O,2015MNRAS.448.3221G,2016ApJ...831...33L,2025ApJ...994...48D,DEBNATH2026100654} of transonic advective viscous disks indicate the presence of thermally driven outflows; however, they do not directly analyze the outflow properties. In this work, we investigate the thermally driven outflows from a transonic advective accretion disk, explicitly accounting for viscous heating. We consider two distinct outer boundary temperatures separated by an order of magnitude. We systematically analyze how this temperature contrast influences the properties of the resulting outflows from the transonic advective disk in the presence of viscous dissipation.

The paper is organized as follows. In Section~\ref{sec:code}, we describe the governing equations and the numerical implementation of the simulation framework. The results are presented and analyzed in Section~\ref{sec:result}. Finally, we summarize our findings in Section~\ref{sec:discussion}.

\section{ Hydrodynamics and Simulation Code} \label{sec:code}
We have considered the same code used in \cite{2025ApJ...994...48D} to solve the conserved fluid equations in the spherical coordinate system $(r, \theta, \phi)$. For completeness, we have briefly mentioned the hydrodynamics equations and code structure here. Assuming the axisymmetry ($\partial/\partial\phi=0$), these equations can be expressed as:
\begin{equation}
 \frac{\partial \mathbf{q}}{\partial t} + \frac{1}{r^2} \frac{\partial (r^2\mathbf{F^r})}{\partial r} +\frac{1}{r \sin\theta} \frac{\partial(\sin\theta \mathbf{F^\theta})}{\partial \theta}  = \mathbf{S}.
 \label{eq:conserve}
\end{equation}
The conserved variables $\mathbf{q}$, corresponding fluxes $\mathbf{F^r}$, $\mathbf{F^{\theta}}$ and the source term $\mathbf{S}$ are given as,
\begin{equation}
\mathbf{q}= 
\begin{bmatrix}
\rho \\
 \rho v_r \\
\rho v_{\theta} \\
 \rho v_{\phi} \\
\frac{\rho v^2}{2} +e \\
\end{bmatrix},
~~ \mathbf{F^r}= 
\begin{bmatrix}
\rho v_r \\
\rho v^2_r +p \\
\rho v_r v_{\theta} \\
\rho v_r v_\phi \\
(E+p)v_r\\
\end{bmatrix},
~~ \mathbf{F^\theta}= 
\begin{bmatrix}
\rho v_\theta \\
 \rho v_\theta  v_r \\
 \rho v^2_{\theta} +p \\
  \rho v_{\theta} v_{\phi}\\
 (E+p)v_\theta \\
\end{bmatrix},
~~ \mathbf{S}= 
\begin{bmatrix}
0 \\
\frac{\rho v^2_{\phi}}{r}+\frac{\rho v^2_{\theta}}{r} -\frac{\rho G M_{BH}}{(r-r_g)^2}+\frac{2p}{r} \\
-\frac{\rho v_r v_{\theta}}{r}+\frac{\rho v^2_{\phi} \cot\theta}{r}+\frac{p \cot\theta}{r}\\
-\frac{\rho v_r v_\phi}{r}-\frac{\rho v_\theta v_\phi \cot\theta}{r}+ S_\lambda \\
 -\frac{G M_{BH} \rho v_r}{(r-r_g)^2}+S_E -S_Q \\
\end{bmatrix}.
\end{equation}
Equation~\eqref{eq:conserve} includes the continuity, momentum, and energy equations. The gravitational field of the black hole is modeled using the Paczy{\'n}sky--Wiita potential $\Phi=-GM_{\rm BH}/(r-\rg)$  \citep{1980A&A....88...23P}. The mass density is denoted by $\rho$, and the total energy density is given by $E=\rho v^2/2+e$, where $e$ is the internal energy density. The pressure is $p$, and the velocity components $(v_r,~ v_\theta,~ v_\phi)$ correspond to the radial, polar, and azimuthal directions, respectively. The specific angular momentum is $\lambda = r v_\phi$, and the total velocity satisfies $v^2 = v_r^2 + v_\theta^2 + v_\phi^2$. We retain only the $r$--$\phi$ component of the viscous stress tensor $W_{r\phi}$.
The viscous stress is given by,
\begin{equation}
W_{r\phi} = \eta_v r \frac{d\Omega}{dr}.
\label{eq:stress}
\end{equation}
where $\Omega$ represents the angular velocity. The dynamic viscosity coefficient is $\eta_v = \rho \nu$,
{$\nu= (\alpha p)/(\rho \Omega_k)$} denotes the kinematic viscosity.
$\alpha$ refers to the Shakura Sunyaev viscosity parameter. $\Omega_k$ is the local Keplerian angular velocity and can be expressed as,
\begin{equation}
\Omega^2_k = -\frac{1}{r} \frac{d \Phi}{d r}.
\label{eq:omegak}
\end{equation}
So, using $W_{r\phi}$ component, $S_{\lambda}$ and $S_E$ are given as,
\begin{equation}
S_{\lambda}= {\frac{1}{r^3}\frac{\partial}{\partial r}\left(r^3 W_{r\phi}\right);} ~
~S_E= \frac{1}{r^2}\frac{\partial ( r^2 v_{\phi} W_{r\phi})}{\partial r},
\label{eq:sesl}
\end{equation}
$S_Q$ is the radiative cooling. We have considered bremsstrahlung and synchrotron radiation. The emissivity resulting from bremsstrahlung (measured in ${\mathrm{erg}~\mathrm{cm}^{-3}~\mathrm{s}^{-1}}$) is given as,
\begin{equation}
Q_{\rm br}=1.4\times10^{-27} n^2_e\sqrt{T_e}\left(1+4.4\times10^{-10}T_e\right).
 \label{eq:bremm}
\end{equation}
and the emissivity due to the synchrotron (in ${\mathrm{erg}~ {\rm cm}^{-3}~ {\rm s}^{-1}}$) is given as,
\begin{equation}
Q_{\rm syn}=\frac{16}{3} \frac{q_e^2}{c} \left( \frac{q_eB}{m_e c} \right)^2 \Theta^2_{e} n_e.
 \label{eq:synchro}
\end{equation}
$T_{\rm e}$ and $n_{\rm e}$ represent the electron temperature and number density, respectively. $\Theta_e=k_b T_e/m_e c^2$ is the dimensionless electron temperature. $q_e$ is the charge of the electron. $B$ represents the random magnetic field, and $B^2/(8 \pi)=\beta p$, where the proportionality parameter $\beta$ is takes as 0.3. So, $S_Q=Q_{\rm br}+Q_{\rm syn}$.

We adopt a variable adiabatic index Equation of State (EoS), as proposed by \cite{2008AIPC.1053..353C,2009ApJ...694..492C}, commonly known as the CR EoS and widely used in numerical simulations \citep{2013ASInC...9...13C,2022ApJ...933...75J,2025ApJ...979...61T,2026Tripathi_et_al}.
The steady-state accretion solution is obtained assuming $\partial/\partial t \equiv 0$, and $v_\theta = 0$. We employ a similar method to get the steady-state solution as in \cite{2024MNRAS.528.3964D}.

The unit of length is the Schwarzschild radius $\rg=2G M_{bh}/c^2$,
the unit of speed is the speed of light $c$, therefore, the unit of time is $t_g=\rg/c$ (where $G$ is the gravitational constant, $M_{bh}$ is the black hole mass). In this system, the unit of specific angular momentum is $\rg c$. Our code is a 2nd-order accurate finite volume code using the HLLC Riemann solver for flux calculation. The code structure, boundary condition, and resolution are kept the same as used in \cite{2025ApJ...994...48D}.

\begin{figure}
\centering
\includegraphics[width=\textwidth]{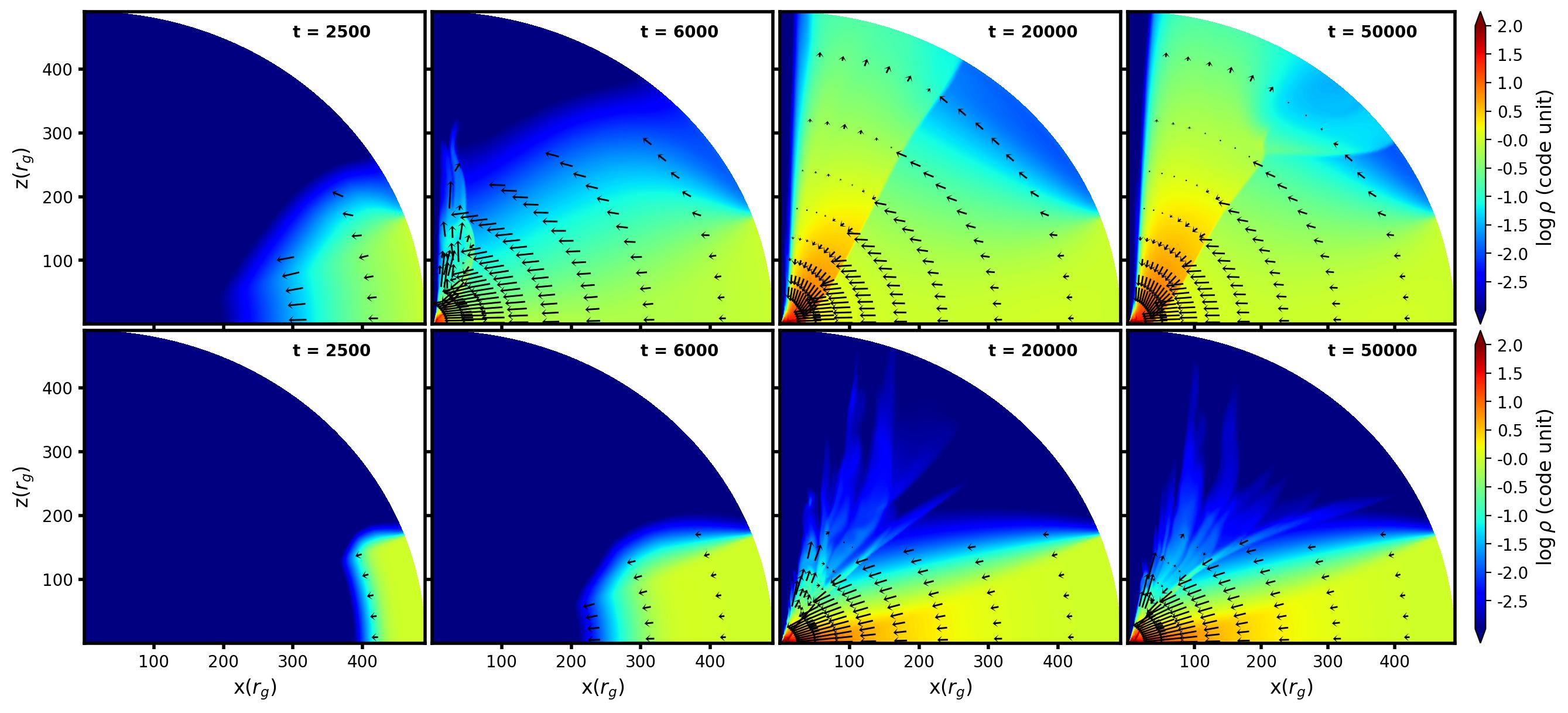}
\begin{minipage}{12cm}
\centering
\caption{\label{fig:1} Snapshots of the density contour with the velocity vector are shown. 1st and 2nd rows are for the model HOT and COLD, respectively. The snapshot time is mentioned in the plots.}
\end{minipage}
\end{figure}

\section{Results} \label{sec:result}
We consider two distinct accretion disk configurations that differ primarily in their thermal properties: a relatively hotter disk and a relatively colder disk. The values of outer boundary radial velocity and temperature are obtained from inviscid analytical solutions for the specific angular momentum $\lambda_{\rm ou}=1.6$ and specific energy $\epsilon=1.0002$ at the fixed radius $r_{\rm ou}=500\,\rg$ (see \cite{2024MNRAS.528.3964D} for details of analytical solution). For this case, the radial velocity is $v_{\rm ou}=-2.095\times10^{-2}$ (in the unit of $c$), the temperature parameter is $\Theta_{\rm ou}=3.585\times10^{-4}$ (corresponding physical temperature is $1.95\times10^9K$) at 500$r_g$, and we refer to this case as model \textit{HOT}. The colder configuration shares the same $v_{\rm ou}$ and $\lambda_{\rm ou}$ but has a lower temperature parameter $\Theta_{\rm ou}=1.195\times10^{-5}$ (corresponding physical temperature is $6.5\times10^7K$); this case is denoted as model \textit{COLD}. Mass accretion rate at the outer boundary is fixed to 0.3 $\medd$, where $\medd=1.44\times10^{17}
(M_{bh}/M_\odot)$gs$^{-1}$.
Matter is injected at the outer boundary with these prescribed values. Snapshots of the density contours overlaid with velocity vectors are presented in Fig.~\ref{fig:1}, where the first and second rows correspond to models \textit{HOT} and \textit{COLD}, respectively. Although both models begin with identical radial velocities, the hotter disk reaches the black hole horizon more rapidly. The \textit{HOT} model subsequently attains a quasi-steady state characterized by the formation of a shock and the emergence of outflows from the post-shock region, whereas the \textit{COLD} model does not exhibit shock formation.
\begin{figure}
\centering
\includegraphics[width=\textwidth]{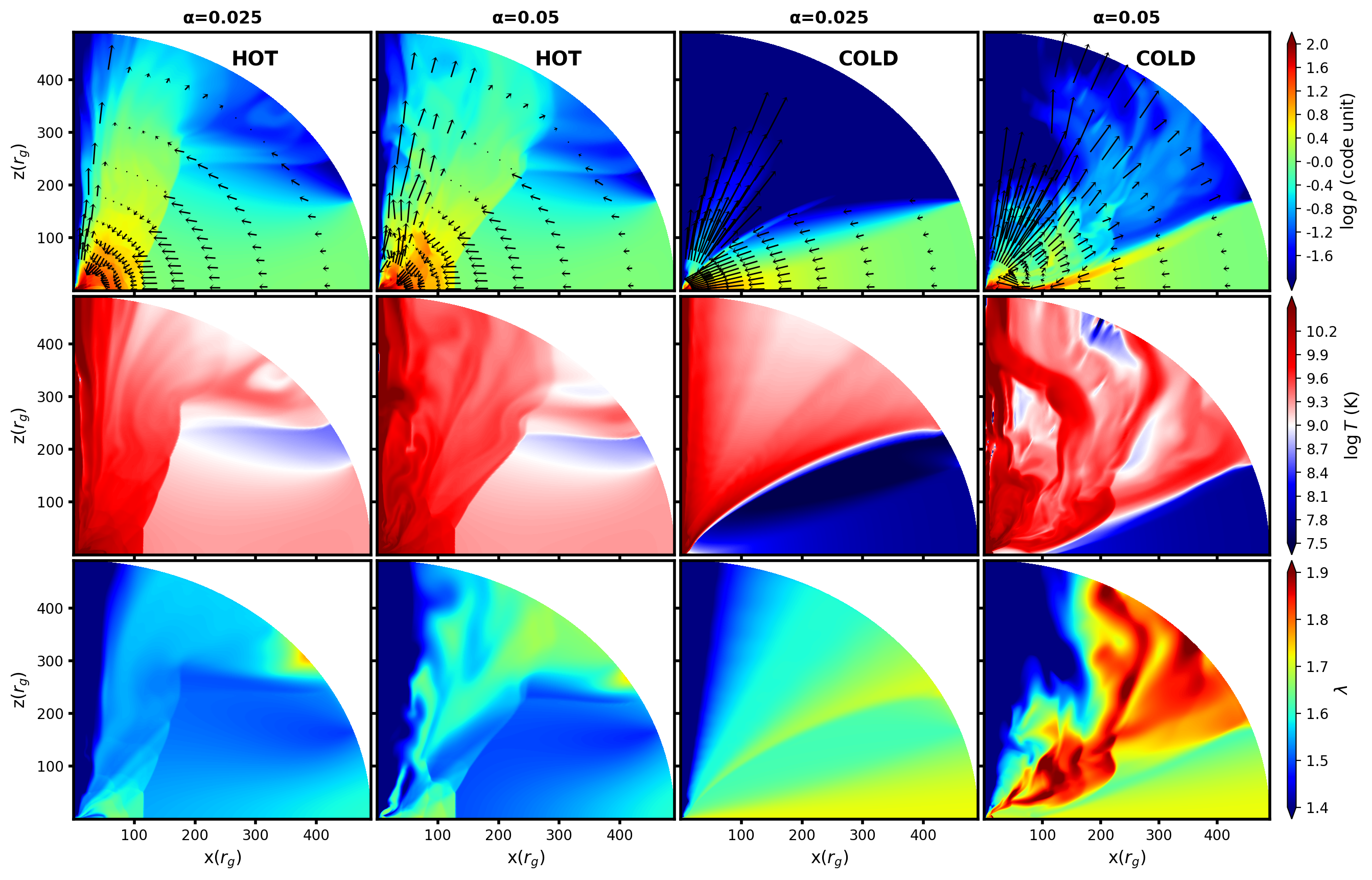}
\begin{minipage}{12cm}
\centering
\caption{\label{fig:2} Snapshots at $t$ = 50000$t_g$ of the density (1st row) with velocity vectors, temperature (2nd row), and angular momentum (3rd row) contour for both the models. The viscosity parameters $\alpha$ are shown in the plots.}
\end{minipage}
\end{figure}
We next incorporate viscosity into the system and evolve the accretion flow further. Figure~\ref{fig:2} shows snapshots of density (1st row) with velocity vectors, temperature (2nd row), and angular momentum (3rd row) contours for both models. Two values of the viscosity parameter, $\alpha=0.025$ and $\alpha=0.05$, are considered for each model. As viscosity is introduced, the strength of the outflow increases due to the enhanced pressure gradients produced by viscous heating. For $\alpha=0.025$, the \textit{COLD} model appears to exhibit weaker outflows compared to the \textit{HOT} model, as is seen from the density contours. However, as $\alpha$ is increased to 0.05, the outflows become stronger in both cases, with the colder disk producing outflows over a wider angular extent. The outflow exhibits a similar temperature distribution in both models. In the HOT model, angular momentum accumulates in the post-shock disk (PSD), while the outflow simultaneously extracts angular momentum from it. In contrast, the COLD model shows a comparatively more efficient removal of angular momentum through the outflow.

\begin{figure}
\centering
\includegraphics[width=0.8\textwidth]{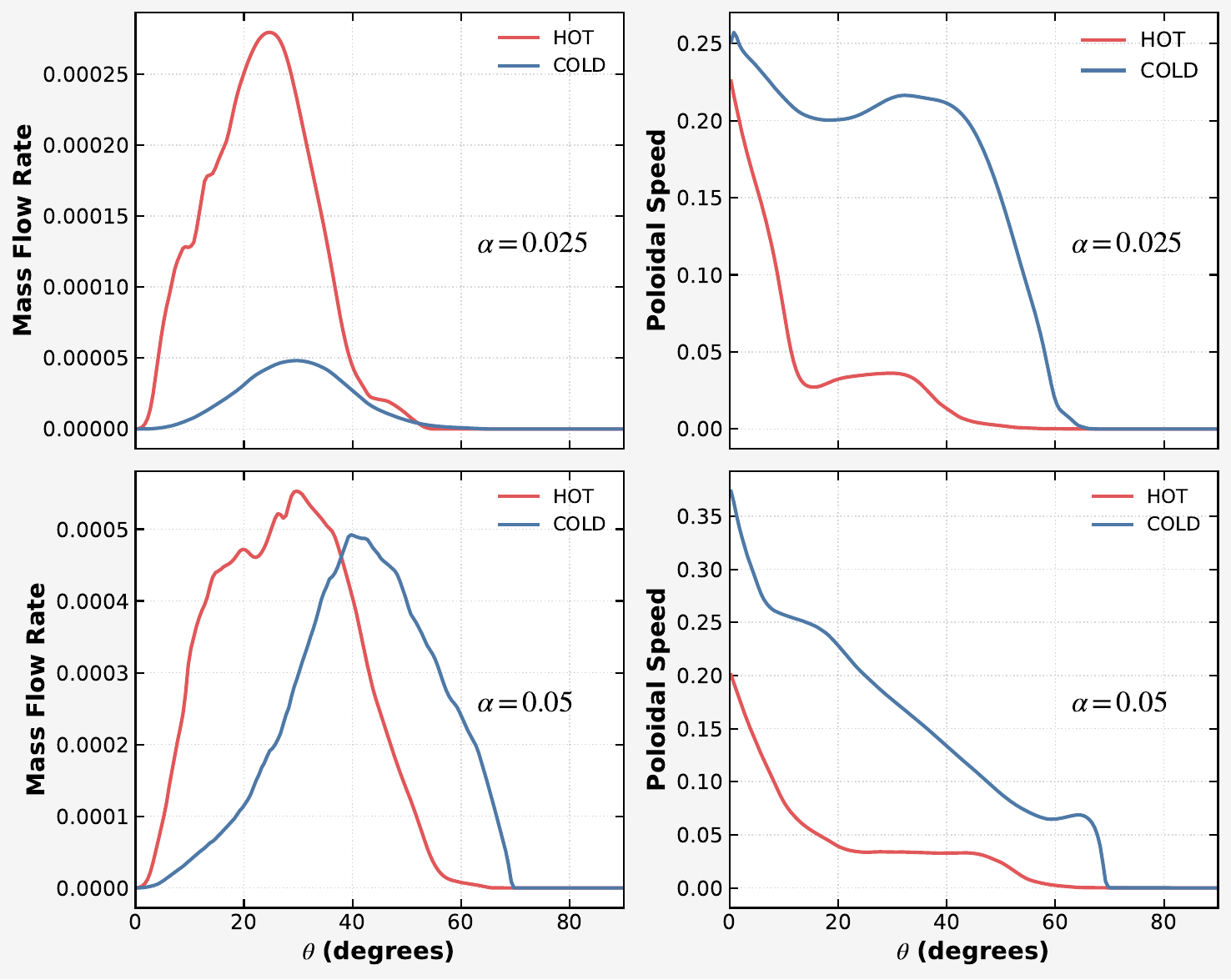}
\begin{minipage}{12cm}
\centering
\caption{\label{fig:3} Time averaged outflow rate (in unit of $\medd$), and the poloidal velocity (in unit of $c$) of model HOT (red) and COLD (blue) for different $\alpha$ = 0.025 (top row) and 0.05 (bottom row).}
\end{minipage}
\end{figure}

Since the instantaneous appearance of outflows in density snapshots doesn't give the overall picture, we quantify the outflow properties more systematically by computing the time-averaged outflow mass flux and the poloidal velocity as functions of the polar angle $\theta$ at the outer boundary of the computational domain. This analysis follows the methodology described in \cite{2025ApJ...994...48D}. Figure~\ref{fig:3} shows the angular distribution of the time-averaged mass outflow rates and poloidal velocity. The total outflow rate and the outflow poloidal velocity increase with increasing $\alpha$, and the maxima shift to higher $\theta$.
\begin{figure}
\centering
\includegraphics[width=0.8\textwidth]{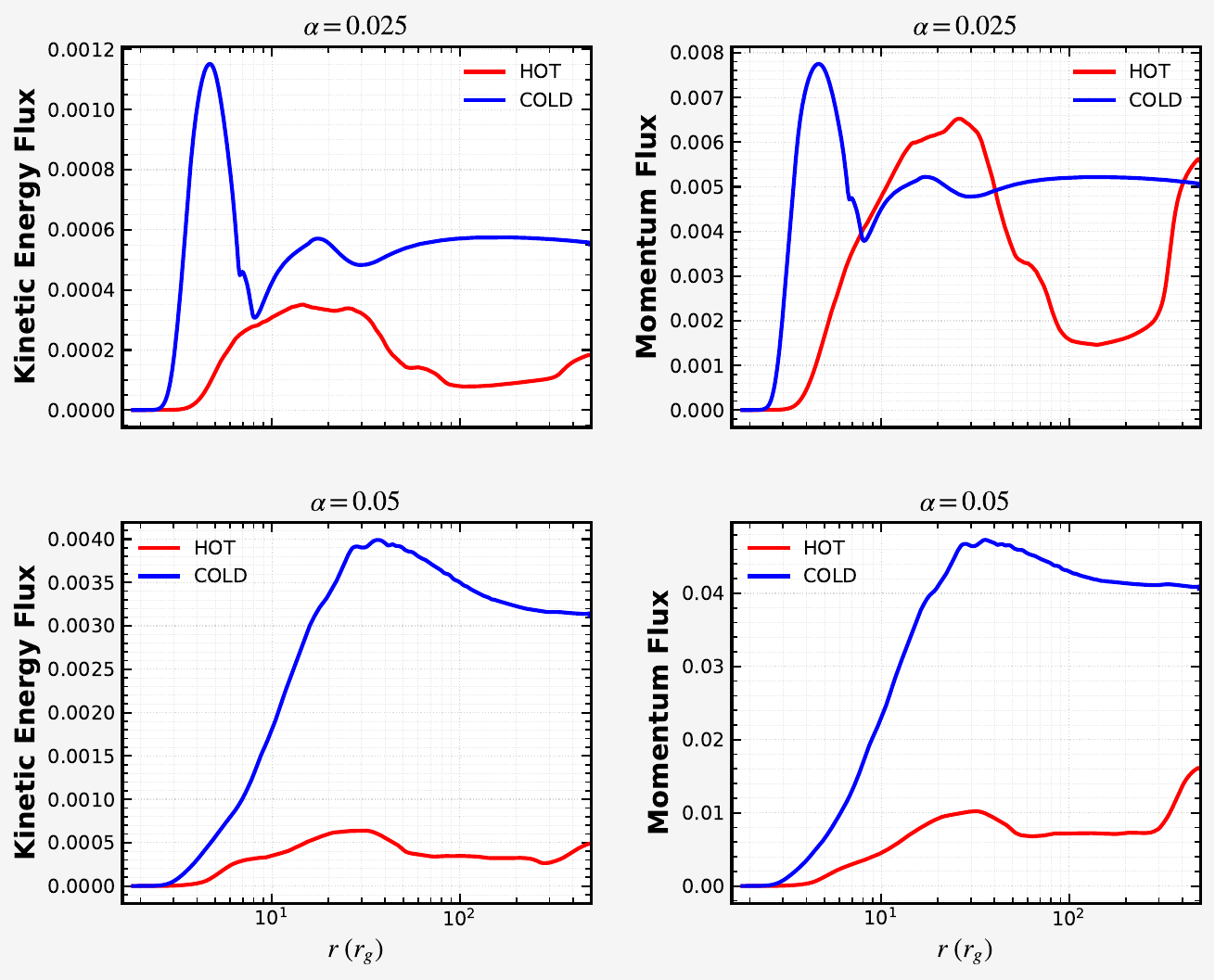}
\begin{minipage}{12cm}
\centering
\caption{\label{fig:4} The time-averaged, $\theta$ integrated, radial distribution of the kinetic energy flux and the poloidal momentum flux of the outflows normalized by $\dot{M}_{\rm acc}c$ and $\dot{M}_{\rm acc}c^2$, respectively.}
\end{minipage}
\end{figure}
For $\alpha=0.025$, the \textit{COLD} model produces a significantly lower time-averaged outflow mass flux at the outer boundary than the \textit{HOT} model, but a much higher poloidal velocity. When the viscosity is increased to $\alpha=0.05$, the outflow rate in the \textit{COLD} model rises substantially and becomes comparable to that of the \textit{HOT} model. Notably, the averaged poloidal velocity close to the axis of symmetry in the \textit{COLD} model exceeds $0.35c$, whereas it remains around $0.2c$ in the \textit{HOT} model. Throughout the computational domain, the \textit{COLD} model consistently shows higher outflow velocities. This behavior is also evident in Fig.~\ref{fig:2}, where the outflowing material in the \textit{COLD} case appears less dense but faster than in the \textit{HOT} case.

To assess the dynamical impact of these outflows on the surrounding medium, we compute the time-averaged and angle-integrated fluxes of kinetic energy and poloidal momentum carried by the outflows. These quantities provide a measure of the ability of the outflows to interact with the ambient gas through energy deposition and momentum transfer. Figure~\ref{fig:4} presents the radial profiles of the kinetic energy flux and poloidal momentum flux for both models with $\alpha=0.025$ and $\alpha=0.05$, normalized by $\dot{M}_{\rm acc}c$ and $\dot{M}_{\rm acc}c^2$, respectively. $\dot{M}_{\rm acc}$ is the accretion rate at the inner boundary. At the outer boundary ($r_{\rm ou}=500\, \rg$ and $\theta=0~\rightarrow~60$ deg), the \textit{COLD} model exhibits a higher kinetic energy flux than the \textit{HOT} model. In addition, the momentum flux in the \textit{COLD} model exceeds $0.04\,\dot{M}_{\rm acc}c$ at this radius, indicating that the outflow is sufficiently powerful to influence the surrounding gas and potentially modify the accretion environment.
Both the kinetic energy flux and momentum flux increase with increasing viscosity. For $\alpha=0.05$, the \textit{COLD} model displays higher fluxes than the \textit{HOT} model across the entire radial range. As shown in Fig.~\ref{fig:4}, cases with higher viscosity exhibit substantially enhanced energy and momentum transport, emphasizing the importance of viscous dissipation in driving strong outflows.
Overall, these results demonstrate that the time-dependent nature of the accretion flow, combined with viscosity-induced heating, is crucial for the generation and maintenance of disk-driven bipolar outflows. These outflows carry significant kinetic energy and momentum, enabling them to interact efficiently with the surrounding medium and potentially influence both the accretion dynamics and feedback processes.

\section{Summary and Discussion} \label{sec:discussion}
We have examined the nature of outflows from transonic advective accretion disks.
A key outcome of this study is that relatively colder transonic advective accretion flows can generate faster and energetically stronger bipolar outflows, although they have a lower mass density, in the presence of viscosity. We theorize that these bipolar outflows, which reach speeds of $>\mbox{few}\times 0.1$c at $500\rg$ would eventually become the astrophysical jets. The viscosity modulates this behavior by enhancing angular momentum transport and increasing the efficiency of matter redistribution within the disk. 

In the \textit{COLD} models, cooling confines the disk into a narrower, conical structure with a relatively cold surface. This geometry creates a high-temperature and high-pressure contrast between the disk surface and the surrounding hot funnel region, producing a strong pressure-gradient force along the disk boundary. The conical shape also provides a larger volume for the outflow to expand into, leading to a broader opening angle, leading to lower mass loading. The reduced inertia of the outflowing material allows the available driving forces to accelerate the gas more efficiently, resulting in higher terminal velocities.
In addition, the outflowing material retains higher specific angular momentum, which enhances centrifugal acceleration and further increases the outflow velocity.
In contrast, the \textit{HOT} models launch denser and more heavily mass-loaded outflows. Although a larger amount of matter is expelled, the available energy is distributed over a greater mass, resulting in comparatively lower velocities. Consequently, the \textit{COLD} disks preferentially produce broad, high-velocity, low-density outflows, whereas the \textit{HOT} disks generate slower but more mass-loaded winds.
Importantly, the higher kinetic energy and momentum fluxes associated with the \textit{COLD} model imply that such outflows can be dynamically more influential, even when their mass flux is comparable to or smaller than that of hotter disks. 

\begin{acknowledgments}
S.D acknowledge the Surya computer cluster facility hosted by ARIES, Nainital, India, and the DANTE platform, APC, Paris, France, for providing computational resources used in this work.
\end{acknowledgments}

\begin{furtherinformation}

\begin{orcids}

  \orcid{0000-0002-9851-8064}{Sanjit }{Debnath}
   \orcid{0000-0002-2133-9324}{Indranil }{Chattopadhyay}
   \orcid{0000-0002-9036-681X}{Raj Kishor}{Joshi}
   \orcid{0000-0001-9094-0335}{Philippe}{Laurent}
   \orcid{0009-0002-7498-6899}{Priyesh Kumar}{Tripathi}
   
\end{orcids}

\begin{authorcontributions}
In this work, the collective efforts were made by all the co-authors with their relevant contributions.
\end{authorcontributions}

\begin{conflictsofinterest}
The authors declare no conflict of interest.
\end{conflictsofinterest}

\end{furtherinformation}



%

\bibliographystyle{bullsrsl-en}

\bibliography{extra}

\end{document}